\documentclass[11pt]{article}
\usepackage{graphicx}
\usepackage{url}
\usepackage{authblk}
\usepackage{booktabs}
\usepackage{tikz}
\usepackage{todonotes}

\usetikzlibrary{arrows.meta, positioning}
\tikzstyle{process} = [rectangle, minimum width=4cm, minimum height=1.2cm, text centered, draw=black, fill=blue!10]
\tikzstyle{startstop} = [ellipse, minimum width=4cm, minimum height=1.2cm, text centered, draw=black, fill=green!20]
\tikzstyle{decision} = [diamond, minimum width=4cm, minimum height=1.5cm, text centered, draw=black, fill=orange!20]
\tikzstyle{arrow} = [thick,->,>=stealth]
\usetikzlibrary{shapes.geometric}

\usepackage{layouts}
\usepackage{subcaption}
\usepackage[hidelinks]{hyperref}

\title{Linking Global Science Funding to Research Publications}

\author[1]{Jacob A. Dalsgaard,}
\author[2]{Filipi N Silva}
\author[3]{Jin Ai}

\affil[1]{Center for Social Data Science, University of Copenhagen, Denmark}
\affil[2]{Indiana University, USA}
\affil[3]{Rutgers University-Newark, USA}

\date{}

\begin{document}
\maketitle

\begin{abstract}
Funding acknowledgments in scholarly publications provide large-scale trace data on organizations that support scientific research. We present a dataset linking global science funding organizations to research publications by systematically disambiguating 7.4 million unique funding acknowledgment strings extracted from publication metadata. Funder names are matched to standardized organizational identifiers using a multi-stage pipeline that combines lexical normalization, similarity-based clustering, rule-based matching, named entity recognition assistance, and manual validation. The resulting dataset links 1.9 million unique funder strings to canonical organization identifiers and records match types and unresolved cases to support transparency. Technical validation includes paper-level comparisons across bibliometric sources and manual verification against full-text acknowledgment sections, with reported recall and precision metrics. This dataset supports analyses of funding flows, institutional funding portfolios, regional representation, and concentration patterns in the global research system.
\end{abstract}

\section{Background \& Summary}
Since Vannevar Bush's postwar vision of publicly supported research, funding has structured the architecture of modern science \cite{bush1945Science}. It underwrites scientific discovery and technological innovation, shaping which problems are pursued, which institutions and regions are resourced, and how research systems evolve over time. Yet the structure and stability of science funding systems are undergoing substantial change worldwide \cite{OECD2025STIOutlook}. Public research funding, historically the primary support for basic and applied science, is increasingly characterized by budgetary volatility, shifting policy priorities, and intensifying cross-sector participation from philanthropic and corporate actors \cite{oecd_msti_highlights_2024, OECD2025STIOutlook}. At the same time, scientific research is becoming more internationally collaborative, with projects, authorship, and funding increasingly spanning national boundaries \cite{nsb_sei_2025_publications}.

Together, these developments heighten the need for integrated and transparent data infrastructures that can systematically link funding information to scientific outputs across countries and sectors. Without such infrastructure, empirical analyses of global science funding remain fragmented and difficult to compare, limiting our ability to trace funding flows, evaluate equity and concentration, or assess how funding systems shape scientific discovery trajectories \cite{kokol_discrepancies_2018}.

However, current empirical analyses of the global research funding landscape remain constrained by fragmentation, inconsistency, and incomplete coverage across existing data sources \cite{kokol_discrepancies_2018}. Funding information is recorded using heterogeneous conventions, extracted from different parts of the scholarly record, and subject to varying disambiguation practices \cite{paul-hus2016Characterization, grassano2017Funding}. These inconsistencies impede interoperability across databases and introduce hidden measurement error into science-of-science and policy research. Moreover, the lack of systematic, paper-level comparisons across major funding databases obscures their relative coverage, strengths, and biases, leaving researchers without clear guidance on how database choice may shape empirical conclusions \cite{kokol_discrepancies_2018}.

Several prominent datasets provide partial but incomplete views of research funding. IRIS UMETRICS offers highly detailed transactional data derived from administrative records at participating U.S. universities, enabling fine-grained analysis of research inputs and expenditures. However, its coverage is largely limited to U.S. institutions and funders \cite{theinstituteforresearchoninnovation&scienceiris2024IRIS}. OpenAlex aggregates funding information from Crossref metadata and applies machine-learning models trained on Microsoft Academic Graph and the Research Organization Registry (ROR) to disambiguate institutions and funders \cite{priem2022OpenAlex, ourresearch2026openalex}. While offering broad and openly accessible coverage, representation of funding sources varies substantially by region and funder type. Dimensions integrates structured grant records with text-mined acknowledgment data and emphasizes grant-level linkage, but exhibits geographic concentration and limited transparency regarding matching algorithms \cite{hook2018Dimensions, kidambi2024grant}. National survey-based datasets such as the Higher Education Research and Development (HERD) Survey and the Business Enterprise Research and Development Survey (BERD) provide valuable aggregate statistics but do not link funding to individual publications, limiting their utility for publication-level funding analysis \cite{herd_survey_2024, brdis_overview}.

A central challenge shared across these infrastructures is the absence of a harmonized, globally comparable mapping of funder identities. Funding acknowledgments are typically recorded as free-text strings, resulting in extensive variation in naming conventions, abbreviations, translations, and levels of organizational granularity. The same funding organization may appear under dozens of name variants, while similar strings may refer to distinct entities. Without systematic disambiguation and crosswalking to persistent identifiers, these inconsistencies reduce reliability, comparability, and reproducibility across studies.

Recent research demonstrates the analytical value of funding acknowledgment data when such challenges are partially addressed. For example, Miao et al. \cite{miao2024Cooperation} use Web of Science (WoS) funding acknowledgments to document patterns of concentration and international interdependence in science funding. At the same time, their findings underscore the sensitivity of results to the completeness and representativeness of underlying funding data. This highlights the need for transparent disambiguation procedures and systematic cross-database validation. This pattern is reinforced in Figure \ref{fig:funder_country}, which shows that the Dimensions database predominantly captures funders from high-income countries, with limited representation across many Global South regions. In contrast, the WoS displays a broader geographic distribution of funding agencies, highlighting the narrower global coverage of funders in the Dimensions.

In this paper, we address these limitations by constructing a large-scale, disambiguated dataset of science funder name strings derived from funding acknowledgments in the WoS and systematically linked, where possible, to canonical organization identifiers from OpenAlex and the Research Organization Registry (ROR). We implement a multi-stage, heuristic-driven disambiguation pipeline that integrates lexical normalization, acronym detection, MinHash locality-sensitive hashing, rule-based matching, named entity recognition assistance, similarity-based fallback methods, and targeted manual validation. Importantly, rather than enforcing exhaustive consolidation, our approach retains unmatched and multiple matched funder strings, thereby preserving transparency about ambiguity and enabling flexible downstream use.

In addition to releasing the dataset, we conduct technical validation and a comparative assessment of funding information across the WoS, OpenAlex, and Dimensions. Using direct paper-level matching, manual validation against full-text funding acknowledgment sections, and comparative analyses of geographic coverage and funder concentration, we quantify differences in recall, precision, and coverage across bibliometric sources. Our results document substantial variation in funder representation, particularly among lower-frequency organizations and funders located in the Global South, underscoring the importance of systematic disambiguation and cross-database benchmarking.

By integrating and crosswalking funder information across major bibliometric databases, this work provides a reproducible data infrastructure for large-scale analysis of the global science funding ecosystem. The resulting dataset enables studies of funding flows across sectors and regions, institutional funding portfolios, concentration and diversity of funders within research domains, and systematic evaluation of biases in existing funding metadata. More broadly, it supports transparent and comparable integration of funding information into science-of-science and policy research.

\section{Methods}

\subsection{Data Sources}

We draw on three major bibliometric databases that include funding information extracted from scholarly publications: WoS, OpenAlex, and Dimensions. Each provides complementary strengths and limitations in coverage, data structure, and approaches to funder disambiguation.

\subsubsection{Web of Science} 
\begin{figure}[ht!]
    \centering
    \includegraphics{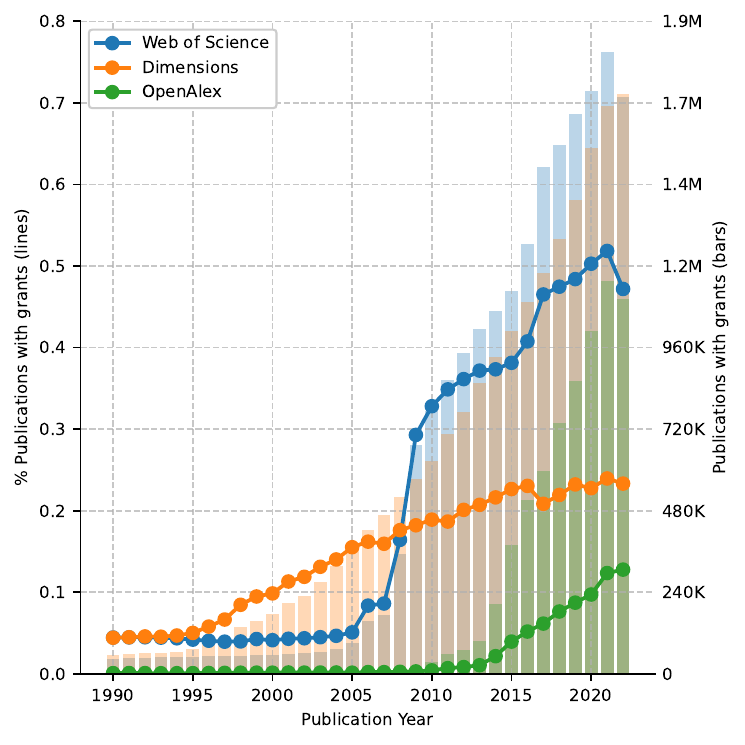}
    \caption{\textbf{Coverage of publications containing funding information across bibliometric databases}.
Left axis shows the yearly proportion of publications with at least one funding acknowledgement (lineplot). Right axis shows the yearly count of publications with at least one funding acknowledgement (bars)}
    \label{fig:coverage}
\end{figure}

Web of Science, developed by Clarivate, is one of the longest-running citation indexing databases, indexing more than 96 million papers. Figure \ref{fig:coverage} shows how from 2005 the coverage of newly published research with funding information rose from 5\% to more than 50\% in 2021. Funding data is recorded via the `Funding Agency` and `Grant Number` fields, extracted through a semi-automated parsing pipeline. Despite known limitations in completeness and standardization, particularly for non-English publications, WoS offers a relatively broad historical coverage and a curated indexing process, making it a widely used source for global funding analyses \cite{paul-hus2016Characterization, grassano2017Funding}. Our pipeline used the WoS CORE-ESCI 2023 snapshot.

\subsubsection{Research Organization Registry} 
The Research Organization Registry is an open, community-governed registry that provides persistent identifiers for research-performing and research-funding organizations worldwide. ROR identifiers are designed to support unambiguous institutional disambiguation across scholarly infrastructures and are increasingly adopted by publishers, funders, and bibliometric databases. As of 2024, ROR contains over 110,000 organizations, including universities, research institutes, hospitals, government agencies, and funding bodies, each represented by a unique identifier, a canonical name, multilingual aliases, geographic metadata, and links to external identifiers such as Wikidata. Our pipeline used ROR v1.72 with the v2 schema \cite{ror2025dataset}.

\subsubsection{OpenAlex} OpenAlex is an open, community-maintained replacement for the Microsoft Academic Graph (MAG), incorporating data from Crossref, institutional identifiers from ROR, and a large-scale classification model to enhance disambiguation of authors and institutions. Until 2025, OpenAlex used metadata from Crossref to infer funding sources when available. In 2026, OpenAlex will start to mine full-text PDFs to match funders to outputs and ingest grant metadata directly from funders \cite{demes2026Funding}. Funders are defined in OpenAlex as organizations that provide financial support for scholarly work. It offers public access via an API and full database snapshots, making it particularly suited for large-scale, reproducible science studies \cite{priem2022OpenAlex}. The latest update date in our OpenAlex snapshot was 2025-09-29.

Institutions in OpenAlex are derived primarily from author affiliations parsed from publications. OpenAlex employs a dual-model classification system, combining a custom deep learning model with a fine-tuned DistilBERT language model, to map noisy, multilingual, and compound institution strings to canonical records. These models are trained using both historical MAG data and synthetic affiliation strings generated from the ROR, enabling broader and more accurate coverage \cite{ourresearch2026openalex}.

\subsubsection{Dimensions} The Dimensions, developed by Digital Science, is a commercial platform that combines publication metadata with grants, patents, and policy documents. Its funding data is derived from multiple sources: direct integrations with funder databases, mining of acknowledgment sections in full texts, and inferred links between grants and outputs. Dimensions emphasizes grant-level linkage, including information about funding agencies, schemes, and award identifiers \cite{kidambi2024grant}. However, due to its proprietary nature, the raw data and matching algorithms are not fully transparent, and access may be restricted \cite{hook2018Dimensions}. We do not use the Dimensions in our pipeline due to restricted access, but in the technical validation, we use a data dump retrieved on January 5, 2025.

\subsection*{Institution Reference Data from OpenAlex and ROR}

To disambiguate the funding organizations listed in the Funding Agency field of WoS, we constructed a unified reference index of organizations using data from OpenAlex and the ROR. OpenAlex provides separate datasets for institutions and funders, but many organizations appear in both roles. We therefore combined these datasets into a single organization list and supplemented missing identifiers and country information using the ROR.

Before merging records, we standardized identifier formats for ROR and Wikidata IDs, normalized organization names to lowercase, and extracted the registered domain from homepage URL to use as a stable web-based name. We then linked records in a stepwise way: first by exact ROR ID, then by exact Wikidata ID, then by shared homepage domain and country.
After this initial linkage, we checked for remaining duplicates (i.e., cases where several records still referred to the same organization). We considered records to represent the same organization when they shared the same normalized name and country. When duplicates were found, we selected a single canonical record. Preference was given to records originating from the OpenAlex funder dataset, followed by those with stable identifiers and more complete activity metadata (e.g., grant counts or publication counts). Since WoS uses ASCII encoding, all names were restricted to ASCII characters.

To retain lexical and search completeness, we merged all known name variants into a single "alternate\_titles" field and stored recognized acronyms both within this field and separately in an acronyms field. For numeric attributes such as grant counts and publication counts, we retained the largest non-missing value across duplicate records to avoid double counting while preserving the most complete information available.

This integrated OpenAlex-based index enables matching against a wide array of entity variants (including acronyms and translated names), facilitating robust disambiguation of funding sources.

\subsection*{Funder Disambiguation and Matching Procedure}

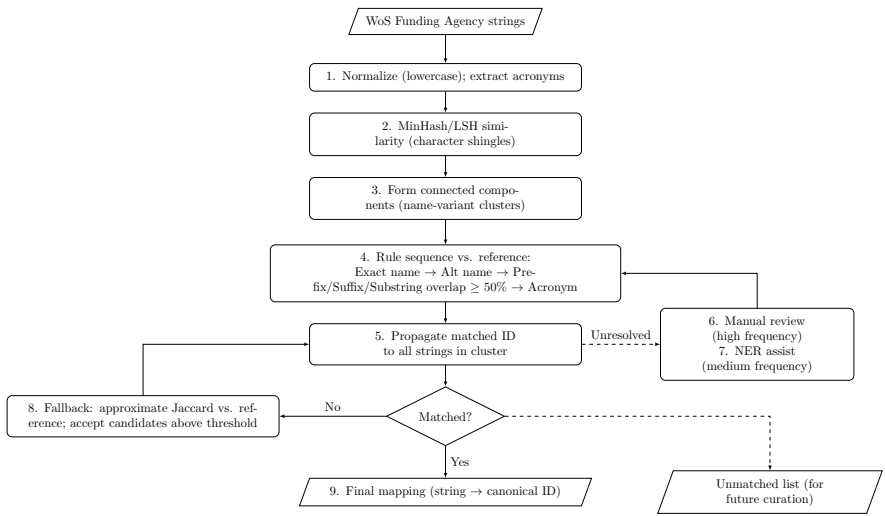
\begin{figure}[h]
\centering
\resizebox{0.92\textwidth}{!}{%
\begin{tikzpicture}[
  node distance=1.1cm and 1.0cm,
  font=\small,
  proc/.style={rectangle, rounded corners, draw, align=center, inner sep=6pt},
  io/.style={trapezium, trapezium left angle=70, trapezium right angle=110, draw, align=center, inner sep=6pt},
  dec/.style={diamond, draw, aspect=2, align=center, inner sep=1.5pt},
  arr/.style={->, >=latex}
]

\node (wos) [io] {WoS Funding Agency strings};

\node (norm) [proc, below=0.8cm of wos, text width=7.2cm] {1. Normalize (lowercase); extract acronyms};
\draw[arr] (wos) -- (norm);

\node (lsh) [proc, below=0.6cm of norm, text width=7.2cm] {2. MinHash/LSH similarity (character shingles)};
\node (cc)  [proc, below=0.6cm of lsh, text width=7.2cm] {3. Form connected components (name-variant clusters)};
\draw[arr] (norm) -- (lsh);
\draw[arr] (lsh) -- (cc);

\node (rules) [proc, below=0.7cm of cc, text width=9.4cm]
{4. Rule sequence vs. reference:\\
Exact name $\rightarrow$ Alt name $\rightarrow$ Prefix/Suffix/Substring overlap $\ge$ 50\% $\rightarrow$ Acronym};
\draw[arr] (cc) -- (rules);

\node (prop)   [proc, below=0.6cm of rules, text width=7.2cm] {5. Propagate matched ID to all strings in cluster};
\node (dmatch) [dec,  below=0.6cm of prop,  text width=2.5cm] {Matched?};
\draw[arr] (rules) -- (prop);
\draw[arr] (prop) -- (dmatch);

\node (fallback) [proc, left=3.0cm of dmatch, text width=7.2cm]
{8. Fallback: approximate Jaccard vs. reference; accept candidates above threshold};
\draw[arr] (dmatch.west) -- node[above]{No} (fallback.east);
\draw[arr] (fallback.north) |- (prop.west);

\node (manual) [proc, right=2.2cm of prop, text width=5.0cm]
{6. Manual review (high frequency)\\ 7. NER assist (medium frequency)};
\draw[arr, dashed] (prop.east) -- node[above]{Unresolved} (manual.west);
\draw[arr] (manual.north) |- (rules.east);

\node (final)     [io, below=0.9cm of dmatch, text width=7.2cm] {9. Final mapping (string $\rightarrow$ canonical ID)};
\node (unmatched) [io, right=2.2cm of final, text width=5.0cm] {Unmatched list (for future curation)};
\draw[arr] (dmatch) -- node[right]{Yes} (final);
\draw[arr, dashed] (dmatch.east) -| (unmatched.north);

\end{tikzpicture}%
}
\caption{\textbf{Overview of the funder name disambiguation pipeline.}
Funding agency name strings extracted from Web of Science acknowledgement metadata are first normalized and clustered using MinHash locality-sensitive hashing to group similar strings. Candidate clusters are then matched to organizations in a reference index derived from OpenAlex and the ROR using a sequence of deterministic rules, including exact-name matching, alternate name matching, substring matching, and acronym matching. High frequency unresolved strings are manually reviewed, and medium-frequency cases are processed using named entity recognition. Remaining unmatched strings are evaluated using a similarity-based fallback procedure before producing the final mapping between funding strings and canonical organization identifiers.}
\label{fig:funder_flow}
\end{figure}

We matched the funder names listed in the Funding Agency field of WoS to the unified organization reference index. The process is illustrated in Figure \ref{fig:funder_flow}.

First, all funder strings were converted to lowercase. Where possible, we also extracted acronyms by considering upper case letters within parentheses (for example, “NSF” from “National Science Foundation (NSF)”)(Step 1). This helped capture organizations that are commonly referred to by abbreviated forms.

Next, to reduce variation caused by formatting differences and minor spelling differences, we grouped similar funder strings together. We measured approximate string similarity using MinHash locality-sensitive hashing and treated strings above a similarity threshold of 0.95 as variants of the same underlying name (Step 2). We constructed a network in which nodes represent funder strings and links connect nodes above the threshold. We grouped the strings by identifying the connected components of the network (Step 3). These groups are treated as candidate clusters representing single organizations.

We then attempted to link each cluster to an organization in the reference index using a structured sequence of matching rules (Step 4). We first checked for exact matches against known names and recorded alternate names. If no exact match was found, we looked for cases where one name was contained within another. We first checked for occurrence as a prefix or suffix before considering any substring matches. In any case, we only considered it a match if the substring covered at least half of the string characters. If still unresolved, we compared acronyms. After matching, the canonical organization identifier was assigned to all names in the cluster (Step 5).

Some strings could not be confidently matched in a single pass through Steps 1-5 based on automated rules alone. We used manual review for the most frequent of these cases, defined as funder strings occurring more than 1,000 times (Step 6). For the manual annotation process, we first searched the OpenAlex API to find candidate matches. Then we manually assigned OpenAlex funder IDs by validating candidates or searching the web. With the manual annotations, we backtracked to Step 4 and applied the manual annotations before running Steps 4 and 5 again. Next, we improved matching for unmatched medium-frequency cases (i.e., funder strings occurring between 100 and 1,000 times) (Step 7). We applied zero-shot named entity recognition using GLiNER \cite{zaratiana2024gliner} to identify organization names within the string before attempting matching again (Steps 4-5). Finally, for any remaining unresolved strings, we performed an additional similarity search by using another LSH index of the remaining strings and querying all possible reference names. We accepted matches above a conservative similarity threshold of 0.9. For details on thresholds, parameters, and models employed, see the Supplementary Information and code.

This approach enables efficient disambiguation across a diverse and multilingual set of funder strings. If any part of the matching process yields multiple plausible candidates, we retain all of them and defer final institution disambiguation to the level of individual papers, where contextual metadata (e.g., co-authors' affiliations, country, subject area) can inform a more precise classification (see Section \ref{resolve}). In particular, our approach prioritizes lexical and structural similarities across strings; this priority is also reflected in the order in which we match the strings. Initial experiments using large language models tended to conflate entities or be confidently incorrect when asked to judge whether matches identified in our pipeline were correct (see Supplementary Information). All steps of the disambiguation pipeline, including preprocessing, clustering, rule-based matching, and manual annotation integration, are implemented in an open-source software package provided with this study. The code repository includes scripts required to reproduce the dataset from the original Web of Science funding strings and the OpenAlex/ROR reference index, subject to the licensing restrictions of the underlying bibliometric databases.

\section{Data Records}

The dataset provides a disambiguated list of global science funders, constructed by matching funder strings from publication acknowledgments to standardized institutional identifiers. The dataset is released as a single Parquet file named \textit{wos\_funders.parquet} and is available from Zenodo \cite{dalsgaard2025dataset}. 

Table \ref{tab:data_fields} describes the data fields associated with the dataset. Each row in the dataset represents a funder string-ID pair. We count the frequency of each funder string and link each string to two standard organization identifiers, the OpenAlex ID and ROR ID, where possible.

\begin{table}[h!]
\centering
\setlength{\tabcolsep}{4pt}
\begin{tabular}{@{}p{0.20\textwidth} p{0.10\textwidth} p{0.20\textwidth} p{0.45\textwidth}@{}}
\toprule
\textbf{Column} & \textbf{Type} & \textbf{\# of Lines with non-empty values} & \textbf{Description} \\
\midrule
\texttt{grant\_agency} & string & 11,086,346 & Original funder name string extracted from publication acknowledgments. \\
\texttt{id} & integer & 5,639,378 & OpenAlex ID for the funder entry. \\
\texttt{counts} & integer & 11,086,346 & Number of publications in which this funder string was identified. \\
\texttt{ids:ror} & string & 5,639,271 & Disambiguated ROR (Research Organization Registry) ID associated with the funder. \\
\texttt{source} & string & 5,639,271 & Whether the ID is an OpenAlex funder ID or an institution ID. \\
\texttt{display\_name} & string & 5,639,271 & Canonical or preferred name for the matched institution from OpenAlex. \\
\texttt{match\_type} & string & 5,639,378 & Specific part of the disambiguation pipeline responsible for this match. \\
\bottomrule
\end{tabular}
\caption{Field descriptions for the disambiguated science funder dataset.}
\label{tab:data_fields}
\end{table}

\section{Technical Validation}
From an initial set of 7.4M unique funder strings, we identify at least one potential match for 1.9M, covering 72\% of all instances of a funder mentioned in a paper in the WoS. In Table \ref{tab:funder_match_quality} we summarize the total and matched number of unique funder strings grouped by how frequent each funder string is. By focusing on structural and lexical similarities the dataset provides good coverage of high and medium frequency funder strings while leaving a large number of diverse and infrequent funder strings unmatched. Table \ref{tab:match_type_counts} describes how many matches each part of the disambiguation pipeline is responsible for.

\begin{table}[ht]
\centering
\begin{tabular}{@{}l r r r@{}}
\toprule
\textbf{Frequency Range} & \textbf{Total Strings} & \textbf{Matched Strings} & \textbf{Unmatched Rate} \\
\midrule
$\geq$ 1000 & 2901    & 2898    & 0.001 \\
100--1000   & 24001   & 19069   & 0.205 \\
10--100     & 263449  & 110836  & 0.579 \\
1--10       & 7108735 & 1881783 & 0.735 \\
\bottomrule
\end{tabular}
\caption{Proportion of unmatched funder strings by frequency range, based on presence of OpenAlex ID.}
\label{tab:funder_match_quality}
\end{table}

\begin{table}[h]
\centering
\begin{tabular}{lr}
\toprule
\textbf{Match Type} & \textbf{Count} \\
\midrule
Not Matched (NaN)     & 5,384,500 \\
Prefix or suffix Match    & 1,049,038 \\
Acronym Match         & 522,745 \\
Substring Match       & 231,125 \\
Document Clustering   & 91,996 \\
Name (Exact)  & 43,021 \\
Alternative names (Exact)    & 42,707 \\
Jaccard Fallback         & 33,307 \\
Manual Annotation     & 647 \\
\bottomrule
\end{tabular}
\caption{Funder string match counts by disambiguation method.}
\label{tab:match_type_counts}
\end{table}

\subsection{Direct matching of papers across datasets}
To assess the recall and complementarity of funder information between the WoS, OpenAlex, and Dimensions, we begin by identifying 2 million articles indexed in both the WoS and OpenAlex and 4 million articles indexed in both the WoS and Dimensions, using the persistent identifier DOI and restricting to records with complete funding information. For each paper, we calculate the intersection of funder IDs in the WoS and OpenAlex. Dimensions uses a different organizational ID (GRID), so we convert between IDs using ROR metadata. Non-empty intersections define a hit. If the intersection returns the full set of funders from either database, we define it as a complete hit. We find that for 88-89\% of papers, at least one funder reported in OpenAlex and Dimensions appears in the WoS, while all funders in 80\% and 67\% of papers in OpenAlex and Dimensions, respectively, are listed in the WoS. Conversely, 66\% and 64\% of papers with funding information in OpenAlex and Dimensions list all funders attributed in the WoS papers (Table \ref{tab:complete_hits_parent}).

\begin{table}[h!]
\centering
\begin{tabular}{lcc}
\toprule
Direction & Complete Hit Rate & Hit Rate \\
\midrule
OpenAlex $\subseteq$ WoS & 0.8 & 0.89 \\
Dimensions $\subseteq$ WoS & 0.67 & 0.88 \\
WoS $\subseteq$ OpenAlex & 0.66 & 0.89 \\
WoS $\subseteq$ Dimensions & 0.64 & 0.88\\
\bottomrule
\end{tabular}
\caption{Directional Intersection and Complete-Hit Coverage Between Databases}
\label{tab:complete_hits_parent}
\end{table}


\subsection{Manual evaluation}
To further assess the accuracy of our disambiguation approach we manually validate a random sample of 250 articles from the WoS by comparing the identified funding agencies against the actual content of the published paper. 
For each paper in the sample:
\begin{enumerate}
  \item We retrieve the full-text version of the paper using its DOI.
  \item We locate and review the funding acknowledgment section manually.
  \item We compare the funder mentions in the paper with the disambiguated set of names from a dataset.
\end{enumerate}

For each paper we record the following:
\begin{itemize}
  \item \textbf{Total number of funders}: We count the total distinct number of funding agencies as reported in the full-text funding and acknowledgment sections. 
  \item \textbf{Number of correct}: We count the number of correctly identified funders
  \item \textbf{Number of incorrect}: We count the number of incorrectly identified funders
\end{itemize}

This labeling allows us to calculate recall and precision for each paper which we summarize using an average over all papers as well as overall statistics by summing over all papers. We additionally report the error rate defined as the number of incorrect divided by the total number of funders of a paper. Finally we report a hit-rate and a all-hits rate defined as the proportion of papers with at least one correctly identified funder and the proportion of papers with all funders identified.

Table \ref{tab:funder_perf} reports the metrics for the WoS, OpenAlex, and Dimensions. The metrics for OpenAlex and Dimensions are based on two subsets of the sample (N=100) with the restriction that each article must be indexed in WoS and Dimensions, and WoS and OpenAlex, respectively, as well as have at least one associated funder. The disambiguated funding information in the WoS generally performs better. Annotations for individual papers are available in the accompanying code repository, along with DOIs and titles.

\begin{table}[h!]
\centering
\begin{tabular}{lccc}
\toprule
 & Avg. Recall & Avg. Precision & Avg. Error Rate \\
\midrule
Web of Science & \textbf{0.78} & \textbf{0.96} & \textbf{0.04} \\
\midrule
OpenAlex & 0.68 & 0.94 & 0.05 \\
\midrule
Dimensions & 0.68 & 0.88 & 0.13 \\
\bottomrule
\end{tabular}

\vspace{0.5cm}
\begin{tabular}{lccccc}
\toprule
 & Recall & Precision & Error Rate & Hit Rate & All-Hits \\
\midrule
Web of Science & \textbf{0.73} & \textbf{0.96} & \textbf{0.03} & \textbf{0.98} & \textbf{0.54} \\
\midrule
OpenAlex & 0.64 & 0.95 & 0.04 & 0.95 & 0.42 \\
\midrule
Dimensions & 0.6 & 0.86 & 0.1 & 0.96 & 0.48\\
\bottomrule
\end{tabular}
\caption{Funder Identification Performance}
\label{tab:funder_perf}
\end{table}

\subsection{Comparative coverage across source databases}
Because the released dataset is constructed from Web of Science funding acknowledgements and evaluated against OpenAlex and Dimensions, it is important to characterize how funding metadata coverage differs across these sources. We therefore compare country-level attribution of funder-linked publications published since 2010 and the concentration of funding records across organizations. These comparisons are intended as technical validation of source-dependent coverage patterns and to document potential biases that users should consider when deciding which funding metadata to use.

\begin{figure}
    \centering

    \begin{subfigure}[b]{\textwidth}
        \includegraphics[width=\textwidth]{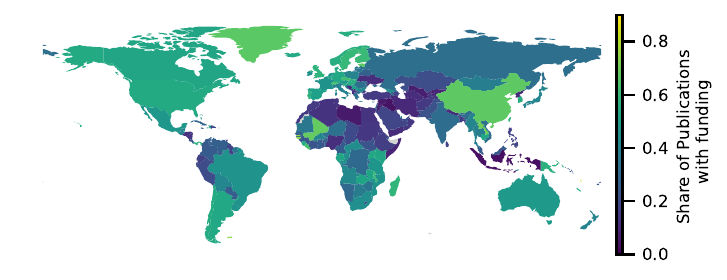}
        \caption{Dimensions}
    \end{subfigure}

    \begin{subfigure}[b]{\textwidth}
        \includegraphics[width=\textwidth]{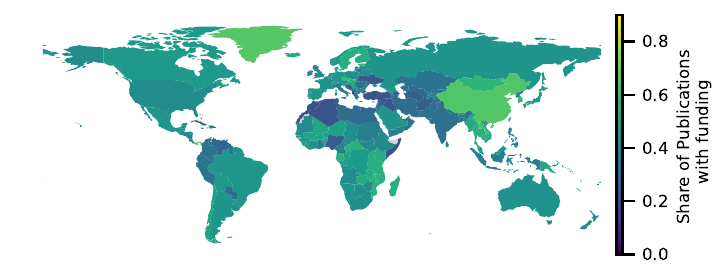}
        \caption{Web of Science}
    \end{subfigure}

    \begin{subfigure}[b]{\textwidth}
        \includegraphics[width=\textwidth]{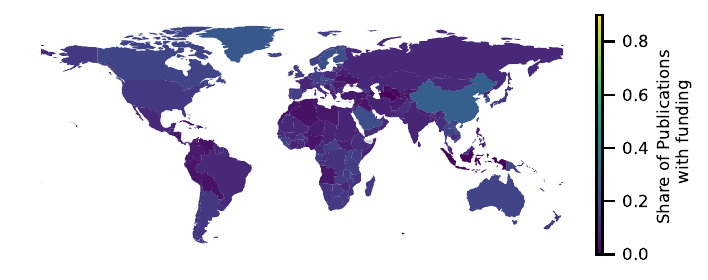}
        \caption{OpenAlex}
    \end{subfigure}

    \caption{\textbf{Country-level coverage of grant attribution by author affiliation.}
Maps show the proportion of publications with recorded funding information by country of author affiliation in (a) Dimensions, (b) Web of Science, and (c) OpenAlex.}
    \label{fig:author_country}
\end{figure}

Figure \ref{fig:author_country} compares the within-country share of publications with recorded funding information by assigning papers to countries through author affiliations, following \cite{miao2024Cooperation}. In this comparison, Web of Science and Dimensions show higher recorded funding coverage in North America, China, and Europe, whereas OpenAlex displays lower coverage overall. Dimensions also shows lower recorded funding coverage across many countries in the Global South. These differences document variation in how funding metadata are captured across sources.

\begin{figure}[htp]
    \centering

    \begin{subfigure}[b]{\textwidth}
        \includegraphics[width=\textwidth]{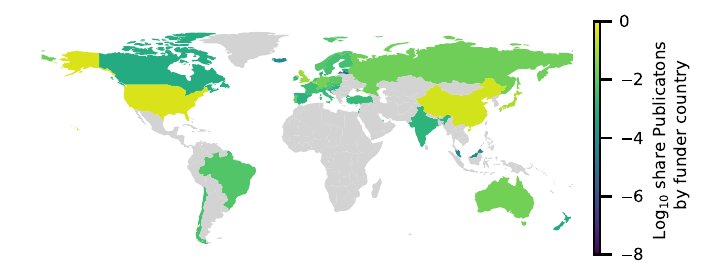}
        \caption{Dimensions}
    \end{subfigure}

    \begin{subfigure}[b]{\textwidth}
    \includegraphics[width=\textwidth]{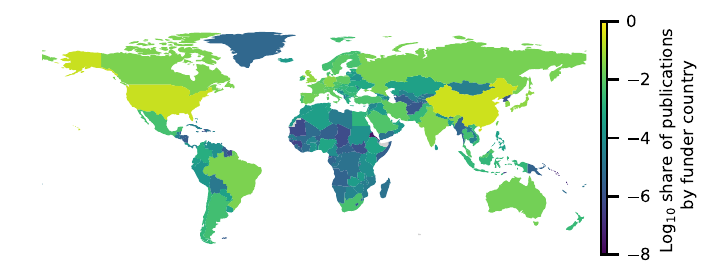}
    \caption{Web of Science}
    \end{subfigure}

    \begin{subfigure}[b]{\textwidth}
    \includegraphics[width=\textwidth]{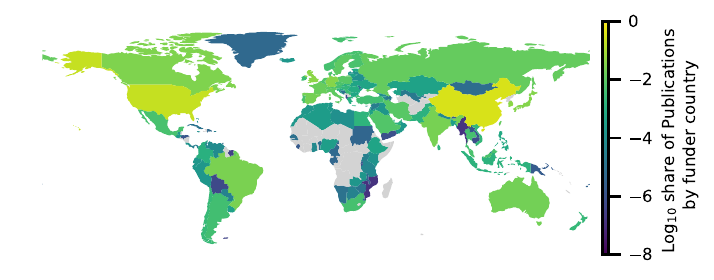}
    \caption{OpenAlex}
    \end{subfigure}

    \caption{\textbf{Geographic distribution of funding agencies across bibliometric databases}.
Maps show the proportion of publications with funding acknowledgements attributed to funders located in each country. Panels compare the geographic distribution of funder-linked publications in (a) Dimensions, (b) Web of Science, and (c) OpenAlex}
    \label{fig:funder_country}
\end{figure}

Figure \ref{fig:funder_country} compares the country distribution of publications with funding information by assigning publications to countries using the locations of matched funding agencies. In this comparison, Dimensions records a larger share of funder-linked publications in high-income countries, while Web of Science displays a broader geographic spread of matched funders, including more funder-linked publications associated with countries in Africa, South America, and Southeast Asia. OpenAlex broadly follows the Web of Science pattern but with visibly sparser coverage in parts of Central and West Africa and Central Asia. These comparisons indicate that the geographic distribution of recorded funders differs substantially across databases and should be taken into account when interpreting country-level analyses.

\begin{figure}[h!]
    \centering
    \includegraphics{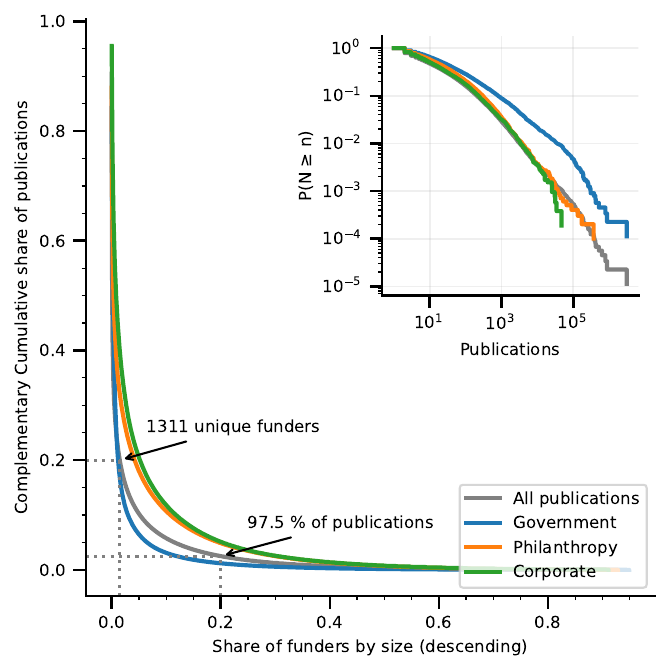}
    \caption{\textbf{Rank–frequency distribution of publications attributed to funding organizations}.
Funding agencies are ranked by the number of publications acknowledging support in the Web of Science dataset. The figure shows the cumulative distribution of publications across funders, illustrating the concentration of funding attribution among a relatively small number of organizations. Sector-specific distributions (government, corporate, and philanthropic) are also shown to highlight differences in the concentration of funding activity across organizational types.}
    \label{fig:rank_fre}
\end{figure}

Finally, figure \ref{fig:rank_fre} presents the rank-frequency distribution of publications. We show that 1311 funders provide funding for 80\% of all publications. While this concentration suggests that focusing on top funders captures most funding activity, it also highlights what might be lost by excluding lower-ranked funders, as 20\% of funders cover 97.5\% of all publications. We also show the concentration of different sectors, showing that governmental funding is concentrated on a small set of institutions compared to corporate and philanthropic funding. Many smaller or regionally specific funding agencies contribute to niche research areas, early-career funding, and diverse institutional support. A dataset like Dimensions, which includes fewer than 700 unique funders, risks overlooking these contributions, potentially biasing analyses toward well-established, high-profile funding organizations.

\section{Usage Notes}
This dataset supports a broad range of analyses at the intersection of science policy, bibliometrics, and innovation studies. Users can pursue diverse inquiries including:

\begin{itemize}
    \item \textbf{Trace funding pathways across sectors and regions.} By linking funder identifiers to institutional metadata, users can study the flow of resources from governmental, philanthropic, and corporate funders to research-producing institutions, facilitating analysis of cross-sectoral collaborations and public-private funding dynamics.
    \item \textbf{Investigate institutional specialization and strategic alignment.} The dataset allows users to map which institutions receive funding from which types of organizations (e.g., national agencies, mission-driven funders, international bodies), enabling studies of institutional portfolios, specialization, and strategic partnerships.
    \item \textbf{Analyze the concentration and diversity of funders in specific research domains.} By combining this dataset with topical classification systems (e.g., field of study, journal categories), one can examine whether research in certain domains is funded by a diverse or narrow set of funders, informing debates about independence, competition, and innovation risk.
    \item \textbf{Geospatial funding flow mapping}: Aggregate funding acknowledgments by country or region of funder origin (via OpenAlex geolocation metadata), enabling spatial visualization of science investment patterns.
\end{itemize}

\subsection{Resolving Ambiguous Matches for Paper-Funder-String Pairs}\label{resolve}

The released dataset provides, for each Web of Science funding agency string, one or more candidate OpenAlex/ROR identifiers. In most cases, a funder string can be mapped to a single canonical organization. However, some strings remain ambiguous and retain multiple plausible candidates. Importantly, this ambiguity occurs at the level of a \emph{paper-funder-string pair}: the same paper may acknowledge several distinct funders, and our goal to resolve which canonical organization corresponds to each acknowledged funder string when multiple candidate matches exist.

To support downstream analyses that require a single identifier per paper-funder-string pair (e.g., aggregating funding by country, sector, or institution), we implement an optional paper-level disambiguation step that selects one candidate organization for each ambiguous pair. This procedure was used in our validation and can be replicated using the accompanying software package.

For any (paper, WoS funder name) pair with multiple candidate matches, we first extract author affiliation locations from the Web of Science metadata. We process authors in positional order: first author, last author, second author, then third author; and attempt to identify the country of each author's affiliation. For each author position, we check whether any candidate funder organization is registered in the same country. If a country match is found for a given author position, we select the matching candidate and terminate.

If no candidate matches any available author-affiliation country, or if multiple candidate funder institutions satisfy the country-matching criterion, we resolve the ambiguity using funder prevalence in OpenAlex. In particular, we select the candidate with the highest overall prevalence in OpenAlex (e.g., frequency of occurrence as a funding organization). This popularity-based criterion serves as a conservative tie-breaker whenever geographic information is unavailable or insufficient to uniquely identify a single institution.

A special case applies to European Union funders. We manually curated a set of EU-level funding organizations (e.g., European Commission, ERC, Horizon programmes). For these entities, we treat an EU-institution candidate as compatible with papers whose authors are affiliated with any country eligible to receive EU funding, including the United Kingdom and associated countries. This exception reflects the transnational nature of EU funding and avoids systematically misclassifying EU funders based on author country alone.

This disambiguation step operates independently for each paper-funder-string pair; thus, after applying it, a single paper may still have multiple funders, one per distinct acknowledged funding agency string.

\subsection{Using Updated Data Sources} The disambiguation pipeline depends on the specific versions of bibliometric and reference data used. Because funding acknowledgement strings and registry records evolve over time, applying the pipeline to updated data releases may produce different string distributions and matching outcomes. In particular, the manual review step (Step 6 in Figure 2), which focuses on high-frequency unmatched strings, is sensitive to the frequency and lexical patterns observed in a given dataset version. Users applying the pipeline to other data snapshots should reassess this step and document any additional modifications to maintain transparency and reproducibility.

\section*{Data Availability}
The disambiguated science funder dataset described in this paper is publicly available on Zenodo \cite{dalsgaard2025dataset}. The dataset is released as a single Parquet file and contains funder name strings extracted from publication acknowledgments together with their matched identifiers. Each record links a funder string to standardized organizational identifiers, including the OpenAlex ID and, where available, the Research Organization Registry ID. The dataset also includes the frequency of each funder string in the Web of Science, the canonical institutional name from OpenAlex, and metadata describing the disambiguation process.

Data from OpenAlex and the Research Organization Registry are released under open licenses (CC0), and their identifiers are included in the dataset in accordance with these licenses. The dataset released with this descriptor contains processed and aggregated information derived from publication acknowledgment fields indexed in the Web of Science. The underlying Web of Science records are proprietary and cannot be redistributed due to licensing restrictions. The shared dataset contains only funder name strings, aggregated counts, and matched organizational identifiers, and does not include any bibliographic metadata that would allow reconstruction of the original database.

\section*{Code Availability}
The code used to perform the disambiguation of funder strings and the paper-level attribution code is available at \href{https://github.com/jacdals97/linking-funding-landscapes/}{https://github.com/jacdals97/linking-funding-landscapes/}. The repository also includes the tables used for manual evaluation.

\bibliographystyle{unsrt} 

\begin{thebibliography}{10}

\bibitem{bush1945Science}
Vannevar Bush.
\newblock Science, the endless frontier: A report to the president on a program for postwar scientific research, 1945.
\newblock NSF reprint, 2023.

\bibitem{OECD2025STIOutlook}
{OECD}.
\newblock {\em OECD Science, Technology and Innovation Outlook 2025: Driving Change in a Shifting Landscape}.
\newblock OECD Publishing, Paris, 2025.

\bibitem{oecd_msti_highlights_2024}
{OECD Directorate for Science, Technology and Innovation}.
\newblock {{OECD}} main science and technology indicators: {{R}}\&{{D}} and related highlights in the march 2024 publication, March 2024.

\bibitem{nsb_sei_2025_publications}
{National Science Board}.
\newblock Discovery: {{R}}\&{{D}} activity and research publications, July 2025.

\bibitem{kokol_discrepancies_2018}
Peter Kokol and Helena~Blažun Vošner.
\newblock Discrepancies among {Scopus}, {Web} of {Science}, and {PubMed} {Coverage} of {Funding} {Information} in {Medical} {Journal} {Articles}.
\newblock {\em Journal of the Medical Library Association : JMLA}, 106(1):81--86, January 2018.

\bibitem{paul-hus2016Characterization}
Ad{\`e}le {Paul-Hus}, Nadine Desrochers, and Rodrigo Costas.
\newblock Characterization, description, and considerations for the use of funding acknowledgement data in {{Web}} of {{Science}}.
\newblock {\em Scientometrics}, 108(1):167--182, July 2016.

\bibitem{grassano2017Funding}
Nicola Grassano, Daniele Rotolo, Joshua Hutton, Fr{\'e}d{\'e}rique Lang, and Michael~M. Hopkins.
\newblock Funding {{Data}} from {{Publication Acknowledgments}}: {{Coverage}}, {{Uses}}, and {{Limitations}}.
\newblock {\em Journal of the Association for Information Science and Technology}, 68(4):999--1017, 2017.

\bibitem{theinstituteforresearchoninnovation&scienceiris2024IRIS}
{The Institute for Research on Innovation \& Science (IRIS)}.
\newblock {{IRIS UMETRICS}} 2024 {{Release Dataset}}, 2024.

\bibitem{priem2022OpenAlex}
Jason Priem, Heather Piwowar, and Richard Orr.
\newblock {{OpenAlex}}: {{A}} fully-open index of scholarly works, authors, venues, institutions, and concepts, June 2022.

\bibitem{ourresearch2026openalex}
OurResearch.
OpenAlex institution parsing.
GitHub https://github.com/ourresearch/openalex-institution-parsing (2026).

\bibitem{hook2018Dimensions}
Daniel~W. Hook, Simon~J. Porter, and Christian Herzog.
\newblock Dimensions: {{Building Context}} for {{Search}} and {{Evaluation}}.
\newblock {\em Frontiers in Research Metrics and Analytics}, 3:23, August 2018.

\bibitem{kidambi2024grant}
Kidambi, M.
Grant data for strengthening research futures.
Dimensions Blog https://www.dimensions.ai/blog/grant-data-for-strengthening-research-futures/ (2024).


\bibitem{herd_survey_2024}
{National Center for Science and Engineering Statistics}.
\newblock Higher education research and development ({{HERD}}) survey 2024, February 2025.

\bibitem{brdis_overview}
{U.S. Census Bureau}.
\newblock Business {R}\&{D} and innovation survey, January 2026.

\bibitem{miao2024Cooperation}
Lili Miao, Vincent Larivi{\`e}re, Feifei Wang, Yong-Yeol Ahn, and Cassidy~R. Sugimoto.
\newblock Cooperation and interdependence in global science funding, February 2024.

\bibitem{ror2025dataset}
Research Organization Registry.
ROR Data (Version 1.72).
Zenodo. https://doi.org/10.5281/zenodo.17280449 (2025).



\bibitem{demes2026Funding}
Kyle Demes.
\newblock Funding metadata in {{OpenAlex}}.
\newblock https://blog.openalex.org/funding-metadata-in-openalex/, January 2026.

\bibitem{zaratiana2024gliner}
Urchade Zaratiana, Nadi Tomeh, Pierre Holat, and Thierry Charnois.
\newblock Gliner: Generalist model for named entity recognition using bidirectional transformer.
\newblock In {\em Proceedings of the 2024 Conference of the North American Chapter of the Association for Computational Linguistics: Human Language Technologies (Volume 1: Long Papers)}, pages 5364--5376, 2024.

\bibitem{dalsgaard2025dataset}
Dalsgaard, J. A., Silva, F. N. \& Ai, J.
Linking global science funding to research publications dataset.
Zenodo. https://doi.org/10.5281/zenodo.18895449 (2025).

\end{thebibliography}

\section*{Author Contributions}
JAI and FNS contributed the initial idea. JAD and FNS are responsible for data collection. The core disambiguation pipeline and and visualizations were developed by JAD who also provided the initial draft of the paper. All authors have contributed to writing the paper. The authors contributed equally to the manual evaluation.

\section*{Competing Interests}
The authors declare no competing interests

\section*{Acknowledgements}
Data sourced from Dimensions, an inter-linked research information system provided by Digital Science \href{https://www.dimensions.ai}{https://www.dimensions.ai}

\section*{Funding}
This work was supported by a research grant (37394) from VILLUM FONDEN. This work utilized Indiana University Jetstream2 CPU through allocation CIS230183 from the Advanced Cyber-infrastructure Coordination Ecosystem: Services \& Support (ACCESS) program, which is supported by National Science Foundation grants \#2138259, \#2138286, \#2138307, \#2137603, and \#2138296. This work was also supported by the National Science Foundation APTO program under grant \#2404109.

\end{document}